\documentclass[aip,amsmath,amssymb,reprint,longbibliography]{revtex4-2}

\usepackage{graphicx}
\usepackage{dcolumn}
\usepackage{bm}
\usepackage[utf8]{inputenc}
\usepackage[T1]{fontenc}
\usepackage{mathptmx}
\usepackage{upgreek}
\usepackage{physics}
\usepackage{siunitx}
\usepackage{layouts}
\usepackage{xcolor}
\usepackage[colorlinks,plainpages=false,linkcolor=blue,urlcolor=blue,citecolor=blue,pdfpagemode=UseNone,pdfstartview=FitH]{hyperref}

\begin{document}

\title[Near transform-limited single photons from rapid-thermal annealed quantum dots]{Near transform-limited single photons from rapid-thermal annealed quantum dots}

\author{H. Mannel}
\email{hendrik.mannel@uni-due.de}
\author{F. Rimek}
\author{M. Zöllner}
\author{N. Schwarz}
\author{F. Schaumburg}
\affiliation{Faculty of Physics and CENIDE, University of Duisburg-Essen, 47057 Duisburg, Germany}

\author{A. D. Wieck}
\author{N. Bart}
\author{A. Ludwig}
\affiliation{Lehrstuhl für Angewandte Festk\"orperphysik, Ruhr-Universit\"at Bochum, 44780 Bochum, Germany}

\author{M. Geller}
\affiliation{Faculty of Physics and CENIDE, University of Duisburg-Essen, 47057 Duisburg, Germany}

\date{\today}

\begin{abstract}
Single-photon emitters are essential components for quantum communication systems, enabling applications such as secure quantum key distribution and the long-term vision of a quantum internet. Among various candidates, self-assembled InAs/GaAs quantum dots (QDs) remain highly promising due to their ability to emit coherent and indistinguishable photons, as well as their compatibility with photonic integration. In this work, we investigate the impact of post-growth rapid thermal annealing (RTA) on the quantum optical properties of single self-assembled QDs embedded in a p-i-n diode structure. The annealing process induces a controlled blueshift of the emission wavelength by promoting Ga in-diffusion and intermixing. Using resonance fluorescence measurements at cryogenic temperatures (4.2 K), we investigate the single-photon statistics, the emission linewidths, and coherence time $T_2$ of the emitted photons. Our results show that, despite the high annealing temperature of \SI{760}{\celsius}, the process does not degrade the optical quality of the quantum dots strongly. Instead, we observe single-photon emission with near transform-limited linewidths, where the dephasing time $T_2$ is only a factor 1.5 above the Fourier-limit $T_2=2T_1$. These findings demonstrate that rapid thermal annealing (RTA) serves as an effective tuning method that preserves the key single-photon emission properties and may help reduce undesirable effects such as non-radiative Auger recombination in quantum photonic applications.
\end{abstract}

\maketitle

\section{Introduction}
Single-photon emitters are one of the fundamental building blocks for quantum communication systems, \cite{Atature.2018} enabling applications such as secure quantum communication \cite{Gisin.2002} or the more long-term vision of a quantum internet.\cite{Kimble.2008, Wehner.2018} A variety of promising single-photon emitters are under investigation,\cite{Aharonovich.2016} ranging from defect centers in diamond \cite{Kurtsiefer.2000,Hepp.2014} and hexagonal boron nitride (hBN)\cite{Tran.2019,Schaumburg.2025} to emerging emitters in other two-dimensional (2D) materials, such as transition metal dichalcogenides (TMDs) \cite{Koperski.2015} and van-der-Waals heterostructure.\cite{Branny.2017,So.2021} However, self-assembled quantum dots (QDs) are still one of the promising candidates with their ability to emit coherent and indistinguishable photons,\cite{Santori.2002,Gazzano.2013,Kuhlmann.2013b} along with their compatibility for integration into photonic circuits.\cite{Lodahl.2015,Schnauber.2021b,Hornung.2024} 

The electronic and optical properties of such quantum dots can be engineered by post-growth treatments such as rapid thermal annealing (RTA).\cite{Sinha2019,Malik.1997,Babinski.2001} This process modifies the dot size and confinement potential by the in-diffusion of Ga atoms inside the QDs,\cite{Keizer2012,Zibik.2007} resulting in an observed blueshift in the excitonic transition energies.\cite{Leon.1996,Manohar.2015} The impact of the RTA process on the optical properties of InAs/GaAs QDs has been explored using techniques such as photoluminescence (PL)\cite{Leon.1996,Manohar.2015} on ensembles of quantum dots and micro-photoluminescence (µPL) on a single InAs QD,\cite{Ellis.2007,Braun.2016}. Resonant excitation, such as four-wave mixing \cite{Langbein.2004} and two-dimensional coherent spectroscopy experiments on RTA-treated quantum dot ensembles \cite{Suzuki.2016,Kosarev.2022}, is rare, and resonant fluorescence on a single self-assembled RTA-treated dot is still missing. However, resonance fluorescence measurements on a single quantum emitter \cite{Muller.2003,Melet.2008,Flagg.2009,Matthiesen.2012,Kurzmann.2019} provide a direct probe of coherent light-matter interactions to observe their spectral linewidth. 

The RTA process is performed at temperatures significantly higher than the decomposition temperature of GaAs. Consequently, it is reasonable to assume that the rapid thermal annealing process may cause damage to the material and degrade the optical quality of the quantum dots. A potential negative impact -- such as additional dephasing of the exciton transition -- can be assessed through resonance fluorescence measurements. In particular, the ratio between the coherence time $T_2$ from the measured linewidth and the radiative lifetime $T_1$ from pulsed resonant measurements, reveals the derivation from the Fourier-transform limit $T_2=2T_1$. Additionally, second-order correlation measurements $g^{(2)}(\tau)$ provide inside into the quality as a single-photon emitter.    

In this work, we present resonance fluorescence measurements of a single self-assembled quantum dot that is embedded in a p-i-n diode structure for electrical charge-state control. The sample was treated after growth by rapid thermal annealing to blue-shift the emission wavelength towards 950 nm. The effects of annealing on the single-photon properties and linewidth are examined and compared to the radiative lifetime $T_1$. The findings demonstrate that post-growth annealing does not have a detrimental effect on the optical properties, and that the near transform-limited photons are observed. 

As previously demonstrated by Gawareck et al.\cite{Gawarecki.2023} for self-assembled quantum dots, changes in the material composition within the dot can significantly influence its physical properties, such as the radiative Auger recombination rate. In colloidal quantum dots (nanoparticles), it has been shown that interfaces play a significant role in modifying or suppressing undesirable effects such as non-radiative Auger recombination\cite{Park.2014}. Controlled alloying has been demonstrated as an effective strategy to tailor these interfaces\cite{Bae.2013b}. Similarly, in self-assembled quantum dots, rapid thermal annealing (RTA) could be employed in the future to tune the emission wavelength and to reduce or suppress detrimental processes such as non-radiative Auger recombination\cite{Kurzmann.2016} or internal photoemission\cite{Lochner.2021}.

The self-assembled quantum dots are embedded in a p-i-n diode structure for electrical control and cooled in a bath cryostat to 4.2 K. The sample was grown using molecular beam epitaxy (MBE). The structure consists of a layer of InGaAs quantum dots in a GaAs matrix embedded in a microcavity (Bragg reflector GaAs/AlGaAs layers) that creates a standing wave electric field with an antinode at the position of the QD layer.\cite{Lochner.2019} The n-doping of the diode acts as an electron reservoir, enabling charging and discharging of the QD via an applied gate voltage $V_g$ and allowing for a tunable shift of the exciton/trion transition energies via the quantum-confined Stark effect. Following the growth phase, the sample was subjected to ex-situ rapid thermal annealing (RTA). The RTA process involved heating the sample to \SI{760}{\celsius} for $30\,$s, with temperatures controlled by a pyrometric measurement on a Si-support wafer, allowing Ga atoms to diffuse into the QDs and promoting intermixing. The annealed sample was then characterized using photoluminescence spectroscopy to analyze its optical properties.

\begin{figure}

    \includegraphics{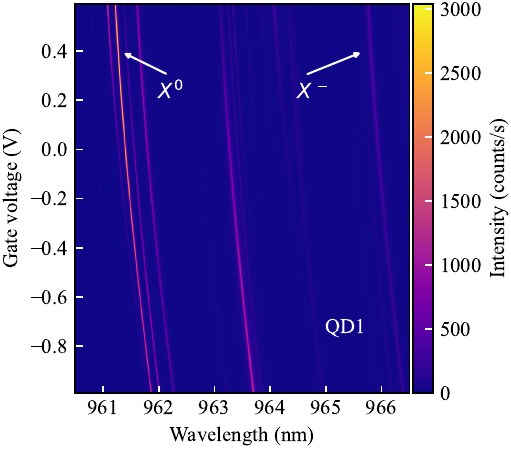}
    \caption{Color-coded photoluminescence (PL) intensity map of the single quantum dot (QD1) as function of the gate voltage and emission wavelength for an excitation laser intensity of $1.78\,\mathrm{\mu W/\mu m^2}$. The color scale represents the PL intensity in counts per second. Two emission lines corresponding to the neutral exciton and the charged trion transition are visible with their voltage-dependent shift by the quantum-confined Stark effect.}
    \label{fig:PL_Scan}
\end{figure}

\begin{figure}
    \includegraphics{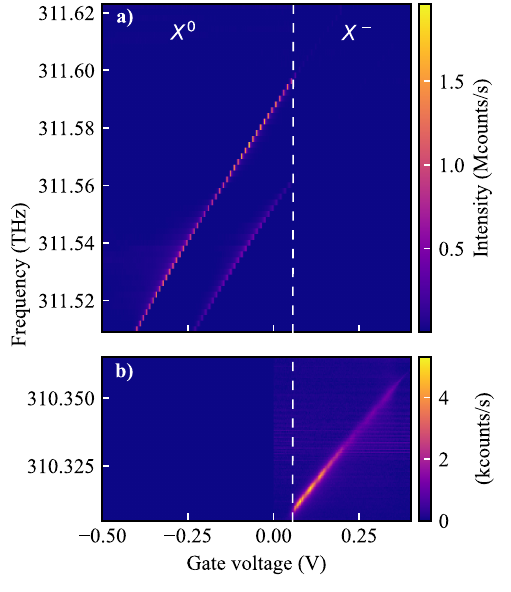}
    \caption{{\bf a)} Resonance fluorescence measurement of the exciton ($X^0$), with its fine structure splitting, and trion ($X^-$) for different laser frequencies and gate voltages. {\bf b)} The trion transition becomes visible for gate voltages above $50$ mV (indicated by the dashed line), where the quantum dot (QD) is charged with a single electron by tunneling from the electron reservoir. Below this voltage, the QD remains uncharged, and at higher frequencies, the exciton transition can be observed.}
    \label{fig:Rf_Scan}
\end{figure}

The initial characterization of the single self-assembled quantum dot (QD1) is shown in Fig.~\ref{fig:PL_Scan} as color-coded micro-photoluminescence ($\mu$-PL) intensity map for different applied gate voltages and an excitation laser intensity of $1.78\,\mathrm{\mu W/\mu m^2}$. The two marked emission lines correspond to the neutral exciton ($X^0$) and negativity-charged trion ($X^1$) transition; confirmed by their energies in the resonance fluorescence measurement in Fig.~\ref{fig:Rf_Scan}. These lines show the well-known quantum-confined Stark shift, while the other lines are due to different charge configurations in the quantum dot.\cite{Ediger.2007} There are no sharp transitions between different charge state configurations at specific gate voltages. Indeed, all excitonic transitions in Fig.\ref{fig:PL_Scan} can be observed over a broad range of applied gate voltages. This behavior results from the weak coupling between the dot and the electron reservoir, caused by a thick tunneling barrier consisting of $10\,$nm AlGaAs and $20\,$nm GaAs. As a consequence, the average electron tunneling rate is low, on the order of 1 $ms^{-1}$.\cite{Kurzmann.2019} and the charge state in the QD is therefore not necessarily in equilibrium with the chemical potential in the back contact if the optical pumping rate exceeds the tunneling rate. As a result, multiple charge states can coexist and are visible in the excitonic transitions at the same gate voltage, as shown in Fig.~\ref{fig:PL_Scan}.

\begin{figure*}
    \includegraphics{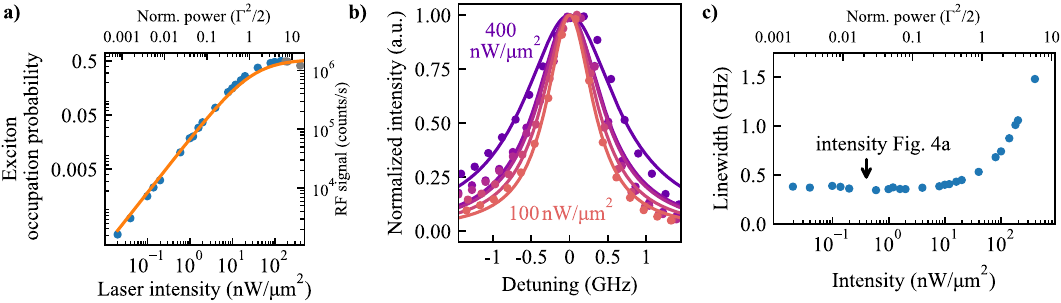}
    \caption{\textbf{a)} Average exciton occupation probability, obtained from integrated photon counts per second for increasing laser excitation intensity. At saturation ($\Omega \approx \Gamma/\sqrt{2}$, with $\Gamma=1/T_1$ and occupation probability equals 0.25) an RF signal of about 1 Mcounts per second is observed. The excitation intensity is also scaled in units of the saturation power $\Omega^2=\Gamma^2/2$. The solid orange line is a fit to the data. \textbf{b)} Power broadening of the exciton transition for selected intensities between 100 $\mathrm{\mu W/\mu m^2}$ and 400 $\mathrm{\mu W/\mu m^2}$. The measured data is shown as dots, while the fits are shown as solid lines. \textbf{c)} Linewidth of the resonant fluorescence from a single quantum dot as a function of excitation intensity. The increase in linewidth at higher intensities is due to power broadening caused by stronger coupling between the excitation field and the dot.}
    \label{fig:fig3}
\end{figure*}

A sharp transition between the exciton and trion transition can be seen in the resonance fluorescence measurement of QD1 in Figure \ref{fig:Rf_Scan}, where the excitonic transitions can only be excited if the quantum dot is in defined charge state, given by the applied gate voltage $V_g$. In Fig.~\ref{fig:Rf_Scan} a) , two distinct spectral lines are now visible that are due to the two bright excition transitions that are separated by the fine-structure splitting, shifting in energy via the quantum-confined Stark effect. A clear charge transition between excition $X$ and trion states $X^-$ occurs at $V_g$= 0.1 V (dashed line). At this voltage, the Fermi level in the electron reservoir is aligned with the s-shell of the dot, allowing an electron to tunnel into the dot. As a result, the trion transition appears at lower energy, as shown in the lower panel, Fig.~\ref{fig:Rf_Scan} b). For the same laser excitation intensity, the resonance fluorescence (RF) intensity of the trion transition is two to three orders of magnitude weaker than of the neutral exciton. This reduction is attributed to non-radiative Auger recombination of the trion state, combined with the weak tunnel coupling to the electron reservoir.\cite{Kurzmann.2016,Lochner.2020} Therefore, the linewidth analysis in the following focuses on the high-intensity neutral exciton transition.

\begin{figure}

    \includegraphics{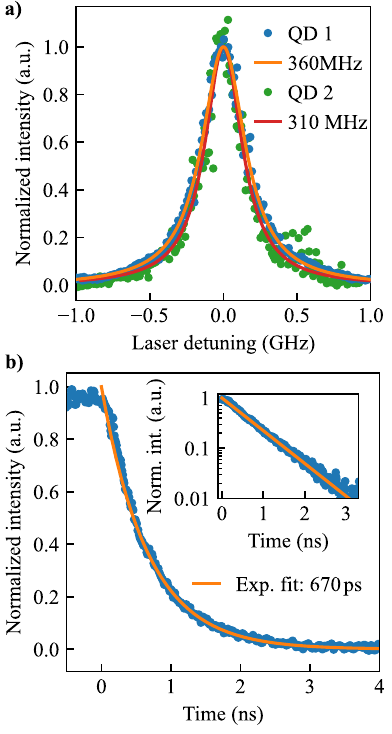}
    \caption{\textbf{a)} Linewidth measurement on two different rapid-thermally annealed quantum dots, QD1 and QD2, receptively. The spectra were obtained for a fixed gate voltage of $V_g= -39.7\,$mV (QD1) and $V_g= 19.5\,$mV (QD2), while a narrow-band single-mode laser (< 1 MHz) is tuned in frequency steps of 10 MHz across the resonances of the exciton transition. A low excitation intensity of $0.5\, \text{nW}/\mu \text{m}^2$ was used to reduce power broadening and the spectra were fitted with a Lorentzian line-shape. \textbf{b)} Time-resolved QD1 fluorescence measurement under resonant pulsed excitation, where an excitation pulse of $1\,$ns was obtained by using an electro-optical modulator. The fit is a convolution of the system response ($270\,$ps) and an exponential decay that yields a $T_1$-time of $670  \pm 30\,$ps. The inset shows the data on a semi-logarithmic scale, demonstrating the single-exponential decay.}
    \label{fig:4}
\end{figure}

Figure \ref{fig:fig3} a) shows a saturation curve of QD1, illustrating the relationship between the laser excitation intensity and the intensity of the emitted QD photons, which corresponds to the average exciton occupation probability. At low excitation intensities, the emission increases linearly as more and more excitons are generated in the quantum dot. With increasing laser power, however, the intensity gradually approches a maximum value ond exhibits characteristic saturation behavior, reaching the average occupation probability of 0.5. As a result of this saturation, power
broadening\cite{Loudon.2010} of the linewidth can be observed in Fig.~\ref{fig:fig3} b) for a laser intensity between 100 and 400 $\mathrm{\mu W/\mu m^2}$. The frequency detuning is obtained here by shifting the QD exciton transition with the quantum-confined Stark effect -- by the applied gate voltage -- in relation to a fixed laser frequency. The linewidth broadening $\Delta \omega_{FWHM}$ (full width half maximum - FWHM) is given by (with $\omega$:\cite{Muller.2007}
\begin{equation}
   \Delta \omega_{FWHM} = \frac{2}{T_2}\sqrt{1+\Omega^2 T_1 T_2}.
    \label{Eq:linewidth}
\end{equation}
Dephasing processes are measured at low excitation intensities, where power broadening is absent. The linewidth is then given directly by the dephasing $T_2$-time by $\Delta \omega_{FWHM}(\Omega=0)=\frac{2}{T_2}$, i.~e.~$\Delta \nu =\frac{1}{\pi T_2}$, which is the same result as the relation between the spectral width and coherence time  for a Lorentzian lineshape; see Saleh et al.\cite{Saleh.2013}. Figure~\ref{fig:fig3} c) shows the measured linewidth as a function of excitation power. At high laser intensities, the linewidth increases to 1.5 GHz due to power broadening. In the low-power regime, a minimum linewidth of $360 \,$MHz is observed for QD1 at an excitation intensity of 0.5$\,\mathrm{n W/\mu m^2}$. The corresponding resonance fluorescence data are shown as blue dots in Fig.~\ref{fig:4} a), where we measured the linewidth with higher accuracy by fixing the gate voltage at $V_g= -39.7\,$mV and tuning the diode laser in frequency steps of $10\,$MHz.  A Lorentzian fit to the data gives a linewidth of 360 MHz (equals 1.49 $\mu$eV and corresponds to $T_2=880$ ps), while a second dot (QD2) shows for comparison an almost identical linewidth of 310 MHz (1.28 $\mu$eV and $T_2=1.03$ ns) for the same excitation intensity, but a gate voltage of $V_g= 19.5\,$mV.

To extract the natural linewidth from the radiative lifetime $T_1$, we employed a pulsed measurement scheme to record the time-resolved decay of the resonance fluorescence (RF) intensity, as shown in Fig.~\ref{fig:4} b) as blue dots. The time-resolved measurements were enabled by short optical pulses generated via an electro-optical modulator (iXblue NIR-MX950-LN) and a pulse pattern generator (Anritsu MP1652A), modulating the continuous-wave diode laser (Toptica CTL950) with a repetition rate of $71\,$MHz and a pulse width of $700\,$ps. The individual RF photons were measured with an avalanche photon diode (ID quantique ID100) and time-correlated with the optical pulse using a time-to-digital converter (Swabian instruments Time Tagger Ultra). The solid orange line represents a fit to the data, yielding a radiative lifetime $T_1$ of $670 \pm 30\,$ps. The fit function is obtained by convolving the system response with a single-exponential decay model corresponding to the quantum dot's spontaneous emission. The system response is primarily determined by the timing jitter of the pulse pattern generator and the avalanche photodiode (APD) and measured to be $270\,$ps. This was characterized in a reference measurement where QD1 was detuned from the excitation laser.

A radiative lifetime of $T_1 = 670\,\mathrm{ps}$ for QD1 corresponds to a Fourier-limited linewidth of $238\,\mathrm{MHz}$, calculated using $\Delta \nu = \frac{1}{2\pi T_1}$.\cite{Saleh.2013}, which equals 0.98 $\mu$eV in energy. For this particular dot QD1, we are by a factor of 1.5 above the Fourier limit, due to additional pure dephasing $T_2^*$. This factor is within the reported values of about 1.4-1.5 of near transform-limited photons from InAs/GaAs quantum dots\cite{Santori.2002,Kuhlmann.2015}, demonstrating the high structural quality of quantum dots exposed to the RTA process. The dephasing time $T_2$ of QD1 obtained from the linewidth measurement in Fig.~\ref{fig:4} a) is $T_2=880$ ps. Together with the radiative lifetime $T_1$ and using $1/T_2 = 1/2T_1 + 1/T_2^*$, we derive a pure dephasing time of $T_2^* = 2.6\,\mathrm{ns}$.    

\begin{figure}
    \includegraphics{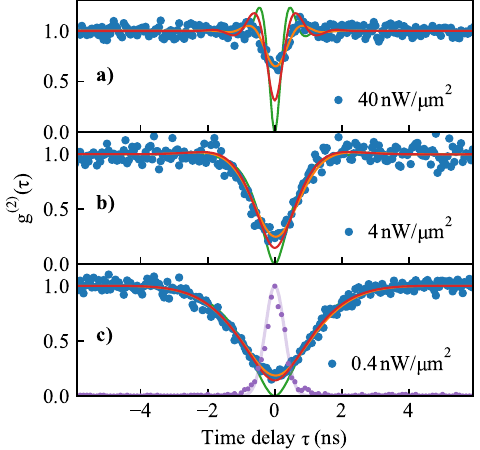}
    \caption{Normalized second-order correlation measurements $g^{(2)}(\tau)$ of QD1 $g^{(2)}(\tau)$ (blue dots) for three different excitation intensities in a) to c). The orange line represents a fit to the data, convolved with the instrument response function (IRF) as purple line. All correlation measurements exhibit pronounced antibunching at zero time delay. The deconvoluted data, shown as solid red lines, yield a second-order correlation function $g^{(2)}(\tau)$ without the instrument response function. The solid green lines represent fits to the data without considering the instrument response function, assuming an ideal single-photon emitter with the parameter A fixed to unity (see text).}
    \label{fig:5}
\end{figure}

Finally, Fig.~\ref{fig:5} shows the normalized second-order correlation function $g^{(2)}(\tau)$ of QD1 for increasing laser excitation intensities from $0.4\, \textrm{nW}/\mu \textrm{m}^2$ to $40\,\text{nW}/\mu \text{m}^2$ intensity, represented as blue dots. The emitted photons were analyzed using a Hanbury-Brown-Twiss-Setup with beam-splitter, two avalanche photodiodes (Excelitas SPCM-AQRH-14-FC) and a time-to-digital converter to measure the $g^{(2)}(\tau)$-function. A pronounced dip at zero time delay ($\tau=0$) is clearly visible in all three measurements, indicating the photon antibunching. The depth of the dip increases with decreasing excitation power, with the strongest antibunching observed at the lowest excitation power, reflecting a minimal probability of multi-photon emission under these conditions. In order to obtain the values of $g^{(2)}(\tau= 0)$, we fitted the experimental data with a convolution of the system response and the theoretically predicted $g^{(2)}(\tau)$-function\cite{Loudon.2010,Scully.1997,Matthiesen.2012}


\begin{equation}\label{eq:g2}
\begin{aligned}
g^{(2)}(\tau) &= 1 - A\, e^{-\eta\,|\tau|}\!\left[\cosh\!\bigl(\lambda\,|\tau|\bigr)+ \frac{\eta}{\lambda}\,\sinh\!\bigl(\lambda\,|\tau|\bigr)\right],\\
\eta &= \frac{3}{4T_1} + \frac{1}{2T_2^*},\\
\lambda &= \sqrt{\frac{1}{4}\!\left(\frac{1}{2T_1} - \frac{1}{T_2^*}\right)^{\!2} - \Omega^2}\,.
\end{aligned}
\end{equation}

For high laser intensities, i.~e.~high Rabi frequencies $\Omega$, the parameter $\lambda$ becomes imaginary. This changes cosh/sinh to cos/sin, enabling Rabi oscillations to be fitted to the measured $g^{(2)}(\tau)$ data. The parameter A accounts for deviations from a perfect single photon emitter.
The background signal was negligibly small and therefore not taken into account in the analysis. The instrument response function (IRF) was measured to be $670\,ps$, shown in Fig.\ref{fig:5} as purple dots with purple lines as fits to the data. The solid orange lines in Fig.~\ref{fig:5} represent fits to the measured $g^{(2)}(\tau)$ data, obtained by convolution of the IRF with the theoretical function in Eq.~\ref{eq:g2} and using the Rabi frequency $\Omega$, the dephasing time $T_2^*$ and $A$ as free fitting parameters. The fits as orange lines show very good agreement with the measured $g^{(2)}(\tau)$ data (blue dots in Fig.~\ref{fig:5}). 

To extract the values of $g^{(2)}(\tau= 0)$ that quantifies the single-photon purity, we evaluated the theoretical function $g^{(2)}(\tau)$ without convolution with the IRF. The solid red lines in Fig.~\ref{fig:5} represent these correlation functions, and the fit parameter $A$ directly corresponds to the single photon purity. The maximum measured value of 86\% for the lowest excitation intensity in Fig.~\ref{fig:5}(c) is a lower bound for the single photon purity of the quantum emitter.



Due to the finite temporal response of the instrument response function (IRF), the measured $g^{(2)}(\tau)$-function is inherently smoothed and broadened by its convolution with the IRF. Consequently, even for an ideal single-photon emitter with a true $g^{(2)}(0) = 0$, the experimentally observed minimum after deconvolution of the IRF remains finite when the IRF width is comparable to, or exceeds, the emitter’s characteristic lifetimes $T_1$ and $T_2$, as in our experiment. When fitting the data, the theoretical function can either be fixed to $g^{(2)}(0) = 0$ (green lines in Fig.~\ref{fig:5}), assuming an ideal single-photon source, or include an amplitude parameter $A$ as a free fitting variable to account for deviations from perfect single-photon emission (red lines in Fig.~\ref{fig:5}). In the latter case, $A$ provides only a lower bound for the single-photon purity. The accuracy of this lower bound could be further improved by enhancing the temporal resolution of the detection system, i.e., by using a narrower instrument response function.

In conclusion, we have investigated the quantum optical properties of single self-assembled InAs/GaAs quantum dots subjected to post-growth rapid thermal annealing (RTA). Despite the high annealing temperature of \SI{760}{\celsius}, we demonstrate that the quantum dots maintain excellent optical quality. Resonance fluorescence measurements reveal linewidths close to the transform limit, with a dephasing time $T_2$ only 1.5 times above the Fourier limit $T_2=2T_1$. Moreover, second-order correlation measurements show single photon emission and a single photon purity at least 86\%, mainly limited by the instrument response function. These findings confirm that RTA can be effectively employed to tune the emission wavelength of InAs quantum dots without compromising coherence or single-photon purity.

The authors have no conflicts to disclose

The data that support the findings of this study are available from the corresponding author upon reasonable request.

\begin{acknowledgments}
	This work was funded by the Deutsche Forschungsgemeinschaft (DFG, German Research Foundation) – Project-ID 278162697 – SFB 1242, and the individual research grant No. GE2141/5-1. A. Lu. acknowledge gratefully support of the DFG by project LU2051/1-1. A. Lu. and A. D. W. acknowledges support by DFG-TRR160, BMBF - QR.X KIS6QK4001, and the DFH/UFA CDFA-05-06. The Mercator Research Center Ruhr (MERCUR) is gratefully acknowledged for support within the project No.~Ko-2022-0013 (H.~M., A.~L., M.~Z.~and M.~G.)
\end{acknowledgments}

\bibliography{Paper-RTA-RF}
	
\end{document}